\documentclass[letterpaper, 10 pt, conference]{ieeeconf}  

\IEEEoverridecommandlockouts                              

\usepackage{graphicx} 
\usepackage{caption}
\usepackage{subcaption}

\usepackage{booktabs}
\usepackage{amsmath}
\usepackage{multirow}
\usepackage{cleveref}

\title{\LARGE \bf
From Flight to Insight: Semantic 3D Reconstruction for Aerial Inspection via Gaussian Splatting and Language-Guided Segmentation
}

\author{Mahmoud Chick Zaouali$^{1}$, Todd Charter$^{1,2}$ and Homayoun Najjaran$^{1, 2}$
\thanks{This work was supported by the National Research Council of Canada (NRC) under the AI4L Supercluster Supplement program.}
\thanks{$^{1}$The authors are with the Faculty of Engineering and Computer Science, University of Victoria, 3800 Finnerty Rd, Victoria, BC, Canada; 
{\tt\small mahmoudchickzaouali@uvic.ca; toddch@uvic.ca; najjaran@uvic.ca}}
\thanks{$^{2}$ Cognia AI Inc.}}

\begin{document}

\maketitle
\thispagestyle{empty}
\pagestyle{empty}


\begin{abstract}
High-fidelity 3D reconstruction is critical for aerial inspection tasks such as infrastructure monitoring, structural assessment, and environmental surveying. While traditional photogrammetry techniques enable geometric modeling, they lack semantic interpretability, limiting their effectiveness for automated inspection workflows. Recent advances in neural rendering and 3D Gaussian Splatting (3DGS) offer efficient, photorealistic reconstructions but similarly lack scene-level understanding.

In this work, we present a UAV-based pipeline that extends Feature-3DGS for language-guided 3D segmentation. We leverage LSeg-based feature fields with CLIP embeddings to generate heatmaps in response to language prompts. These are thresholded to produce rough segmentations, and the highest-scoring point is then used as a prompt to SAM or SAM2 for refined 2D segmentation on novel view renderings. Our results highlight the strengths and limitations of various feature field backbones (CLIP-LSeg, SAM, SAM2) in capturing meaningful structure in large-scale outdoor environments. We demonstrate that this hybrid approach enables flexible, language-driven interaction with photorealistic 3D reconstructions, opening new possibilities for semantic aerial inspection and scene understanding.
\end{abstract}


\section{INTRODUCTION}

As aerial inspection is increasingly applied to infrastructure and environmental monitoring, augmenting visual data with spatial structure through 3D reconstruction offers new possibilities for automation and analysis. While 2D imagery remains the foundation of most UAV-based inspection workflows, it provides limited spatial understanding, making it difficult to assess geometric relationships or identify subtle structural anomalies.

Traditional photogrammetry methods, including Structure-from-Motion (SfM) \cite{snavely2006photo, schoenberger2016sfm} and Multi-View Stereo (MVS) \cite{1640800}, produce dense geometry from UAV imagery but lack semantic awareness and struggle to scale efficiently. Their reliance on feature matching across many high-resolution images makes them computationally intensive and error-prone in large or repetitive scenes.

Neural rendering techniques like Neural Radiance Fields (NeRF) \cite{mildenhall2021nerf} offer improved visual quality from sparse views, but are limited by high training cost and poor scalability. Recently, 3D Gaussian Splatting (3DGS) \cite{kerbl20233d} has emerged as a faster, photorealistic alternative, representing scenes as anisotropic Gaussians that support efficient training and real-time rendering. However, 3DGS remains purely geometric, lacking the semantic capabilities needed for inspection tasks such as defect detection or object-level querying.

To address this, recent work incorporates features from vision-language foundation models like CLIP \cite{radford2021learning, cherti2023reproducible}, LSeg \cite{li2022languagedriven}, and SAM \cite{kirillov2023segany, ravi2024sam2} into 3DGS, enabling language-guided interaction with reconstructed scenes. These methods show promise for tasks like segmentation and open-vocabulary grounding, but have primarily been tested in indoor or controlled environments.

Outdoor aerial scenes pose unique challenges: variable lighting, large spatial scales, and visually repetitive or textureless surfaces make semantic reconstruction more difficult. Existing language-embedded 3DGS pipelines have not been evaluated in these settings.

In this work, we adapt Feature-3DGS \cite{zhou2024feature} to UAV-based outdoor inspection and propose a novel, two-stage segmentation pipeline. We first generate CLIP-LSeg heatmaps in response to language prompts, then use the peak activation as a point prompt for SAM or SAM2 to refine segmentation. We evaluate this pipeline qualitatively on two UAV-captured datasets and demonstrate its feasibility for open-vocabulary, language-driven interaction with large-scale 3D scenes.

Our contributions are:
\begin{itemize}
    \item We adapt Feature-3DGS to outdoor aerial inspection and compare semantic feature extractors (CLIP-LSeg, SAM, SAM2).
    \item We propose a language-guided segmentation pipeline combining CLIP-based heatmaps with point-prompted SAM refinement.
    \item We provide qualitative results demonstrating the strengths and limitations of foundation models for language-based 3D scene understanding in UAV scenarios.
\end{itemize}

The remainder of this paper is organized as follows: Section~\ref{sec:background} reviews related work. Section~\ref{sec:methodology} describes our pipeline. Section~\ref{sec:results} presents experimental findings. Section~\ref{sec:conclusion} concludes the paper and outlines future directions.

\section{BACKGROUND}\label{sec:background}

Understanding the limitations of traditional aerial reconstruction pipelines and the opportunities introduced by neural scene representations requires reviewing several key areas. This section first provides an overview of UAV-based aerial inspection and conventional photogrammetry methods. We then discuss recent advances in 3D scene representation, with a focus on 3D Gaussian Splatting. Next, we review vision-language foundation models and their role in bridging visual and semantic understanding. Finally, we examine recent efforts to embed language features into 3D reconstructions, which form the basis for the methods adapted in this work.

\subsection{Aerial Inspection/Photogrammetry}

Unmanned aerial vehicles (UAVs) have become a key platform for infrastructure monitoring, land surveying, and disaster assessment due to their flexibility, high-resolution imaging capabilities, and ability to access difficult environments. Traditional aerial inspection workflows rely heavily on photogrammetry pipelines, such as Structure-from-Motion (SfM) and Multi-View Stereo (MVS), to generate 3D reconstructions from overlapping images. These techniques reconstruct geometry by detecting and matching image features across views, triangulating sparse point clouds, and densifying them into surface models.
While photogrammetry has enabled widespread adoption of UAV-based 3D mapping \cite{isprs-archives-XXXVIII-1-C22-25-2011}, it faces notable limitations: reconstructions often lack semantic understanding, suffer from loss of fine-grained detail, and scale poorly with increasing dataset sizes. Processing large outdoor scenes requires extensive computational resources, and traditional pipelines are prone to alignment errors, particularly when scenes are divided into smaller chunks for processing. As inspection tasks demand higher fidelity, faster turnaround, and contextual scene understanding, there is growing interest in augmenting or replacing conventional photogrammetry with neural and hybrid 3D scene representations.

\subsection{Scene Representation and 3D Gaussian Spaltting}

Recent advances in 3D scene representation have introduced neural approaches that aim to improve upon traditional geometry-based methods. Neural Radiance Fields (NeRF) \cite{mildenhall2021nerf} model volumetric scenes as continuous functions, achieving highly realistic renderings from sparse views by learning dense 3D radiance fields. Despite their success, NeRF-based methods are computationally intensive, requiring long training times and significant memory consumption, which limits their applicability for large-scale inspection tasks.
3D Gaussian Splatting (3DGS) \cite{kerbl20233d} offers a faster and more scalable alternative, representing scenes as collections of anisotropic Gaussian primitives that can be directly splatted onto image planes during rendering. This representation enables efficient training and real-time rendering while maintaining photorealistic quality. 3DGS models are particularly attractive for UAV-based inspection, where rapid reconstruction and visualization are critical. However, like most neural scene representations, standard 3DGS captures only photometric and geometric information, lacking semantic interpretability necessary for downstream inspection tasks such as defect detection, classification, or semantic querying.

\subsection{Vision-Language Foundation Models}

Vision-language foundation models (VLMs) align visual and textual modalities by learning shared embedding spaces from large-scale image-text datasets. Models such as CLIP and ALIGN \cite{pmlr-v139-jia21b} demonstrate strong zero-shot generalization, enabling open-vocabulary classification, retrieval, and grounding without task-specific fine-tuning. Their semantic flexibility has made them valuable across tasks including object detection, segmentation, and captioning. In 3D vision, CLIP-based embeddings have been used to enrich reconstructions with semantic meaning, supporting tasks like language-guided segmentation and scene annotation.

Concurrently, vision foundation models (VFMs) have advanced general-purpose visual understanding without language supervision. Transformer-based architectures like ViT \cite{dosovitskiy2020image} and Swin \cite{liu2021swin}, along with self-supervised models such as MAE \cite{he2022masked}, BEIT \cite{bao2021beit}, and DINO \cite{caron2021emerging}, have enabled scalable learning from unlabeled data. DINOv2 \cite{oquab2023dinov2} further improved generalization by training large ViT backbones on curated datasets. The Segment Anything Model (SAM) introduced prompt-based zero-shot segmentation, supporting flexible image annotation across diverse domains. Recent efforts combine multiple foundation models (e.g., SAM with Grounding-DINO and DeAOT) \cite{ren2024grounded} to tackle complex tasks like interactive video segmentation. Extensions to 3D, including SAM-driven point cloud segmentation \cite{zhou2024point}, highlight growing interest in transferring 2D semantic priors to 3D scene understanding.

\begin{figure*}[ht]
    \centering
    \includegraphics[width=\linewidth]{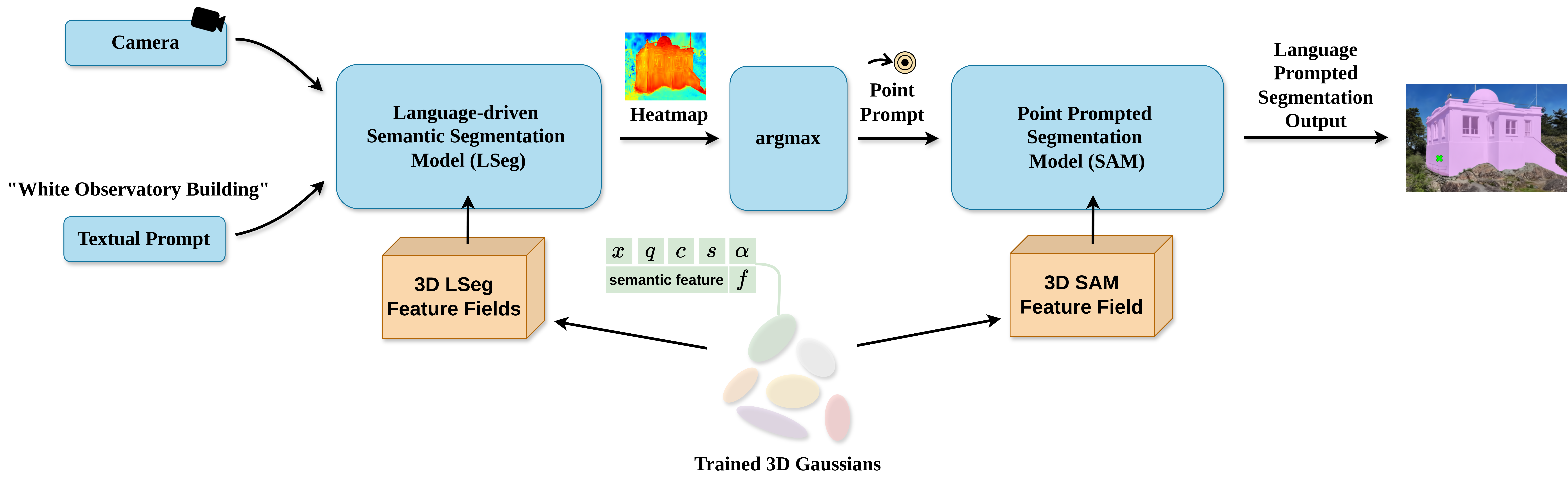}
    \caption{End-to-End Pipeline for Language-Guided 3D Reconstruction and Semantic Feature Field Distillation}
    \label{fig:feature-3DGS}
    \vspace{-1em}
\end{figure*}

\subsection{Language-Embedded 3D Feature Fields}
To incorporate semantic understanding into 3D scene representations, recent research has focused on embedding language-driven features directly into 3D reconstruction pipelines. LangSplat \cite{qin2024langsplat} proposed extending 3D Gaussian Splatting by distilling features from pre-trained vision-language models into each Gaussian primitive, enabling 3D visual grounding and dense captioning tasks. By associating semantic embeddings with localized 3D elements, LangSplat facilitates open-vocabulary querying and contextual scene interpretation.
Feature-3DGS \cite{zhou2024feature} further advanced this direction by refining feature distillation strategies to improve spatial precision and semantic coherence, enabling richer feature fields within 3DGS frameworks. These approaches demonstrate the potential of augmenting photorealistic reconstructions with dense semantic information, but to date, have primarily focused on indoor or small-scale environments where lighting conditions, object scales, and scene compositions are relatively constrained. Extending these methods to outdoor aerial inspection tasks introduces new challenges related to scale variance, environmental complexity, and dynamic lighting, motivating the adaptations proposed in this work.

\section{METHODOLOGY} \label{sec:methodology}

\subsection{UAV Data Collection and Preprocessing}
To evaluate Feature-3DGS in outdoor aerial inspection, we collected two UAV datasets of infrastructure scenes. Images were captured at 1080p resolution using a DJI Mini Pro 4 drone with significant overlap and varying angles. COLMAP was used for structure-from-motion and sparse point cloud generation. Gaussian scene primitives were initialized using the standard 3DGS preprocessing pipeline.

\subsection{Feature-3DGS with Visual and Language Feature Fields}
We train separate Feature-3DGS models using feature fields from three sources: LSeg (CLIP-based), SAM1, and SAM2. The standard dual-branch Feature-3DGS optimization is applied, with photometric loss for appearance and a distillation loss for feature field supervision. Training is performed for 7k iterations on each model.

\subsection{Language-Guided Segmentation Pipeline}
We extend the LSeg-trained Feature-3DGS with language-guided segmentation. Given a text prompt, we compute a cosine similarity heatmap between the prompt and CLIP-embedded feature fields rendered from the scene. A threshold is applied to obtain a rough binary segmentation. The point with highest heatmap activation is then used as a prompt to SAM or SAM2, yielding a refined segmentation on the same 2D novel view. An overview of this methodology is presented in \Cref{fig:feature-3DGS}.

\subsection{Training and Implementation Details}
All models are implemented in PyTorch with customized cuda kernels for per-pixel Gaussians feature field rasterization and trained on an NVIDIA RTX Titan GPU with 24GB VRAM. Training follows the learning schedules provided in the original Feature-3DGS repository. We use pre-trained checkpoints from CLIP, SAM1, SAM2, and LSeg for feature extraction.

\subsection{Datasets and Evaluation Setup}

We evaluate our Feature-3DGS-based pipeline on two custom UAV-captured datasets designed for outdoor aerial inspection:

\textbf{Building dataset:} A single-story structure in an urban park, surrounded by paved paths, vegetation, and a picnic area. It includes windows, doors, and roof features. The dataset comprises 223 high-resolution images (195 training, 28 test) captured at varying altitudes and angles.

\textbf{Observatory dataset:} A dome-roofed observatory surrounded by open landscape, with structural elements such as staircases, railings, and rooftop equipment. The dataset contains 253 images (221 training, 32 test) with full multi-view coverage.

Camera poses were estimated using COLMAP \cite{schoenberger2016sfm}, and 3D Gaussian Splatting primitives were initialized using the recovered parameters. Feature fields were rendered from novel viewpoints for qualitative analysis and semantic querying experiments.

\section{EXPERIMENTAL RESULTS} \label{sec:results}

We evaluate our pipeline on the two datasets representing outdoor aerial inspection scenarios: a small public building and a white observatory structure. 3D reconstructions were generated using the training images for each dataset, and image-based rendering metrics were computed on novel views rendered from the test set camera parameters. We report peak signal-to-noise ratio (PSNR), structural similarity index measure (SSIM), and learned perceptual image patch similarity (LPIPS). These metrics, shown in \Cref{tab:metrics}, indicate that our reconstructions achieve high visual fidelity, consistent with prior results on indoor and synthetic scenes.

\begin{table}[h]
\centering
\caption{Feature 3DGS reconstruction metrics across different scenes.}
\label{tab:metrics}
\begin{tabular}{l|ccc}
\toprule
\textbf{Metrics} & PSNR$\uparrow$ & SSIM$\uparrow$ & LPIPS$\downarrow$ \\
\midrule
Building    & 22.0039 & 0.6406 & 0.3958 \\
Observatory & 21.8115 & 0.6778 & 0.3386 \\
\bottomrule
\end{tabular}
\end{table}

\begin{figure*}[h]
     \centering
     \begin{subfigure}[b]{0.195\textwidth}
         \centering
         \includegraphics[width=\textwidth]{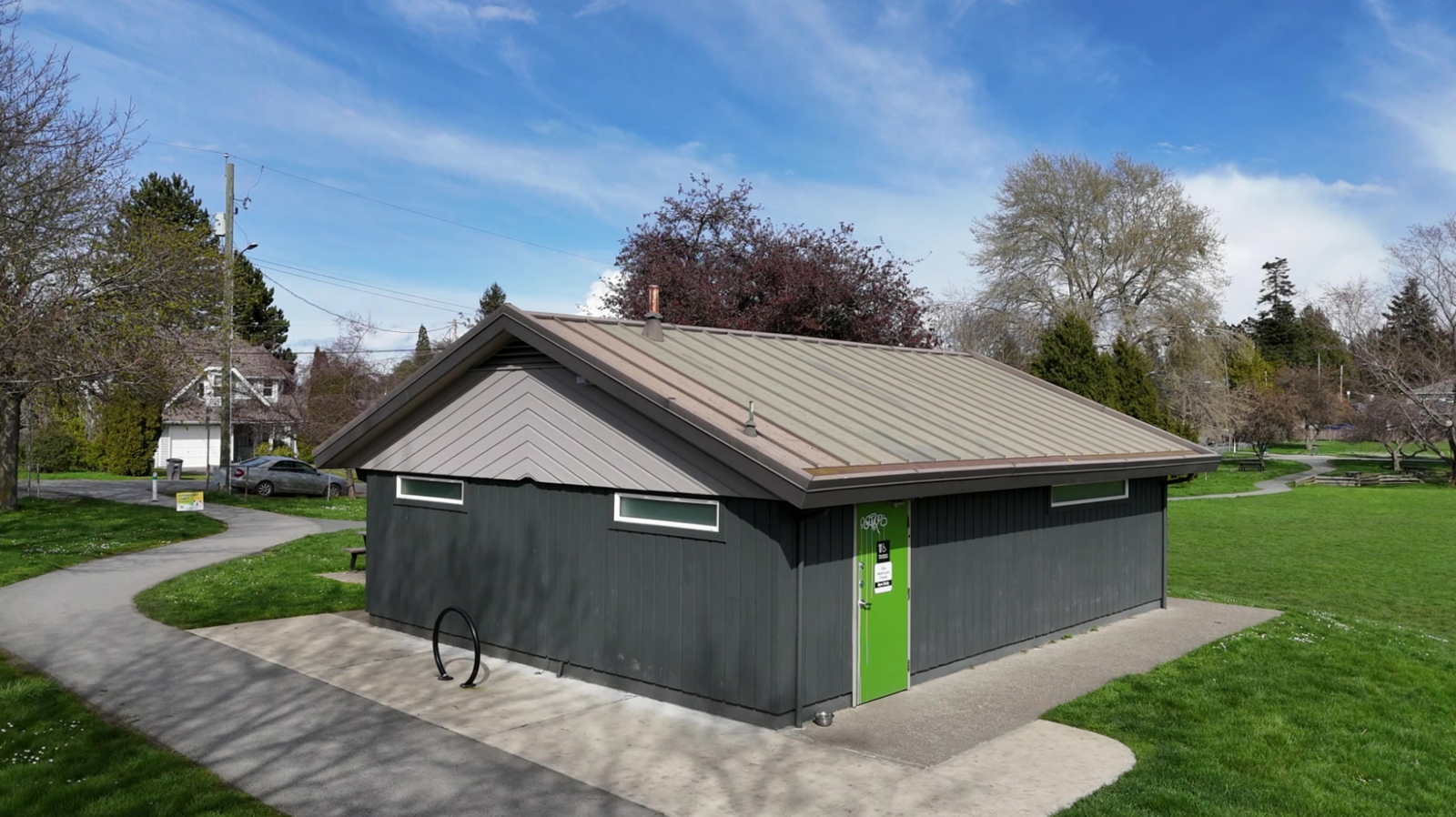}
         \caption{Ground Truth Image}
         \label{FFB-GT}
     \end{subfigure}
     \hfill
     \begin{subfigure}[b]{0.195\textwidth}
         \centering
         \includegraphics[width=\textwidth]{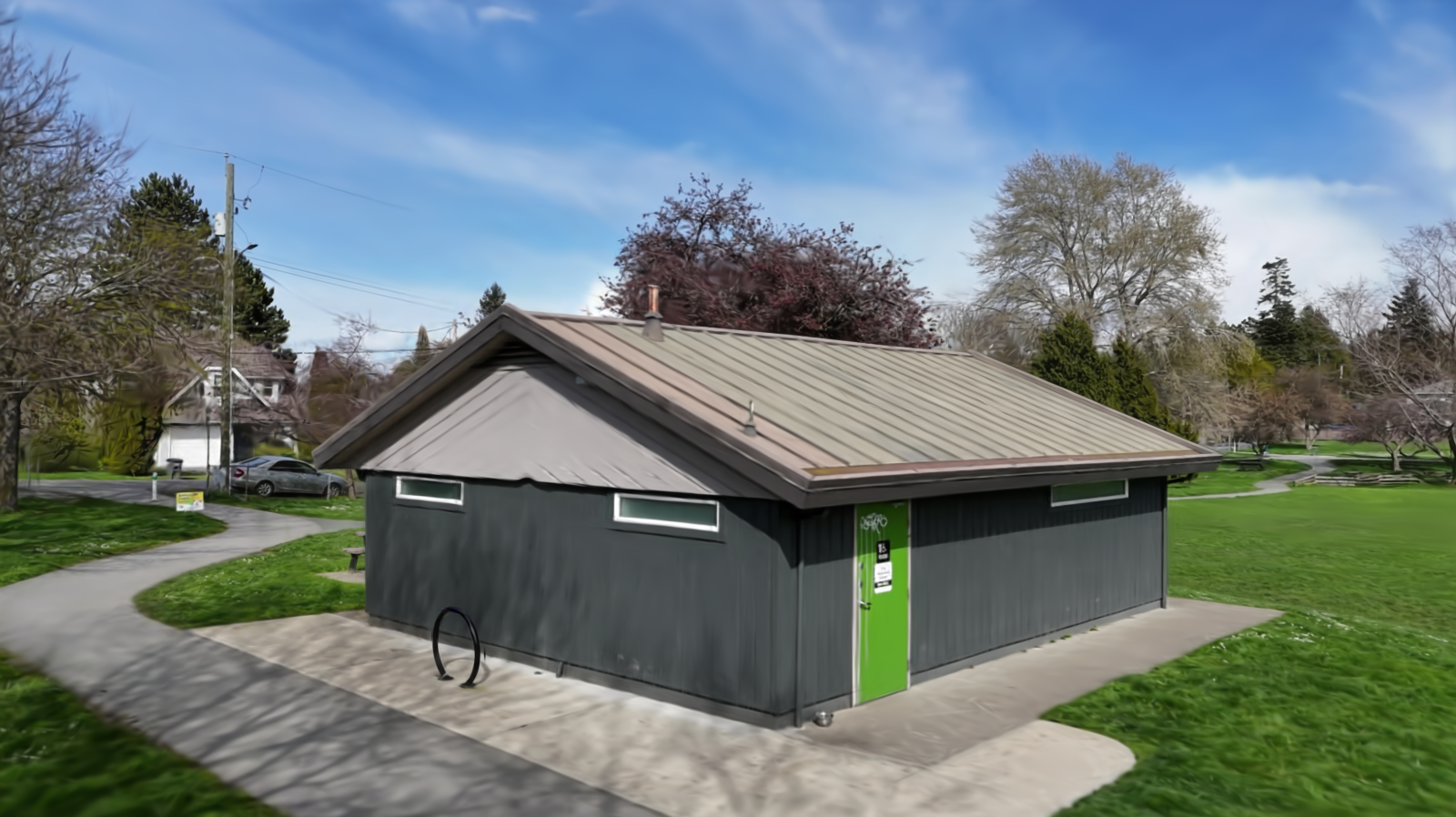}
         \caption{3DGS Rendering}
         \label{FFB-GS}
     \end{subfigure}
     \hfill
     \begin{subfigure}[b]{0.195\textwidth}
         \centering
         \includegraphics[width=\textwidth]{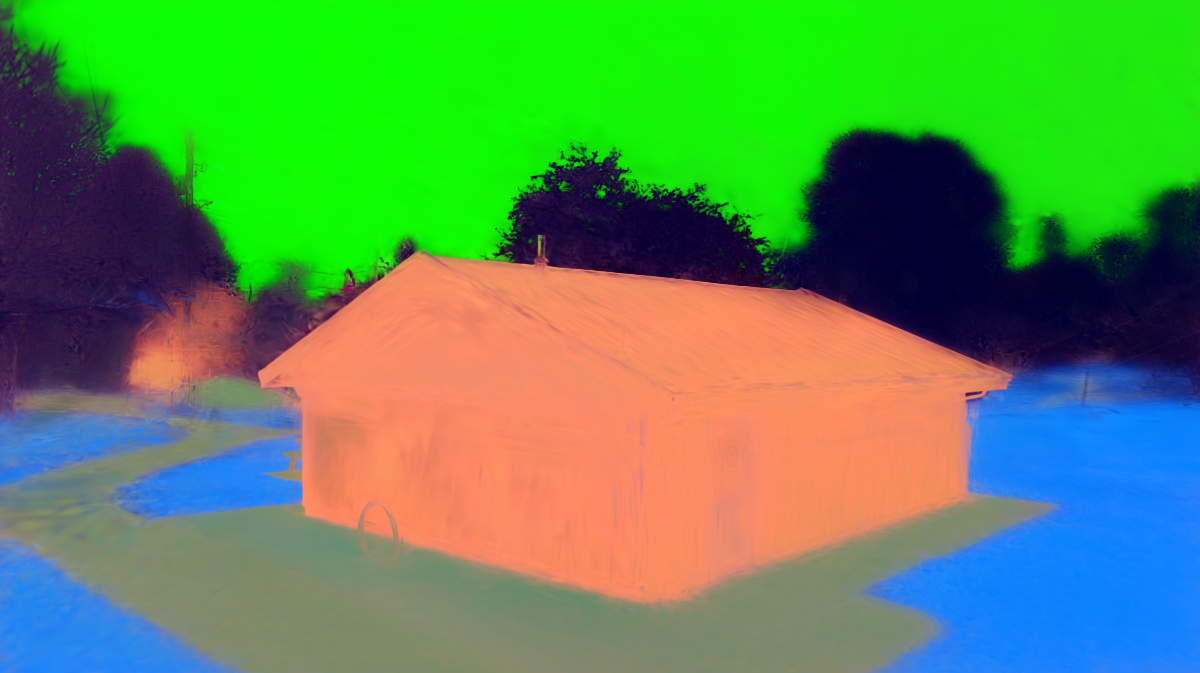}
         \caption{LSeg Feature Field}
         \label{FFB-Feature-3DGS-LSeg}
     \end{subfigure}
     \hfill
     \begin{subfigure}[b]{0.195\textwidth}
         \centering
         \includegraphics[width=\textwidth]{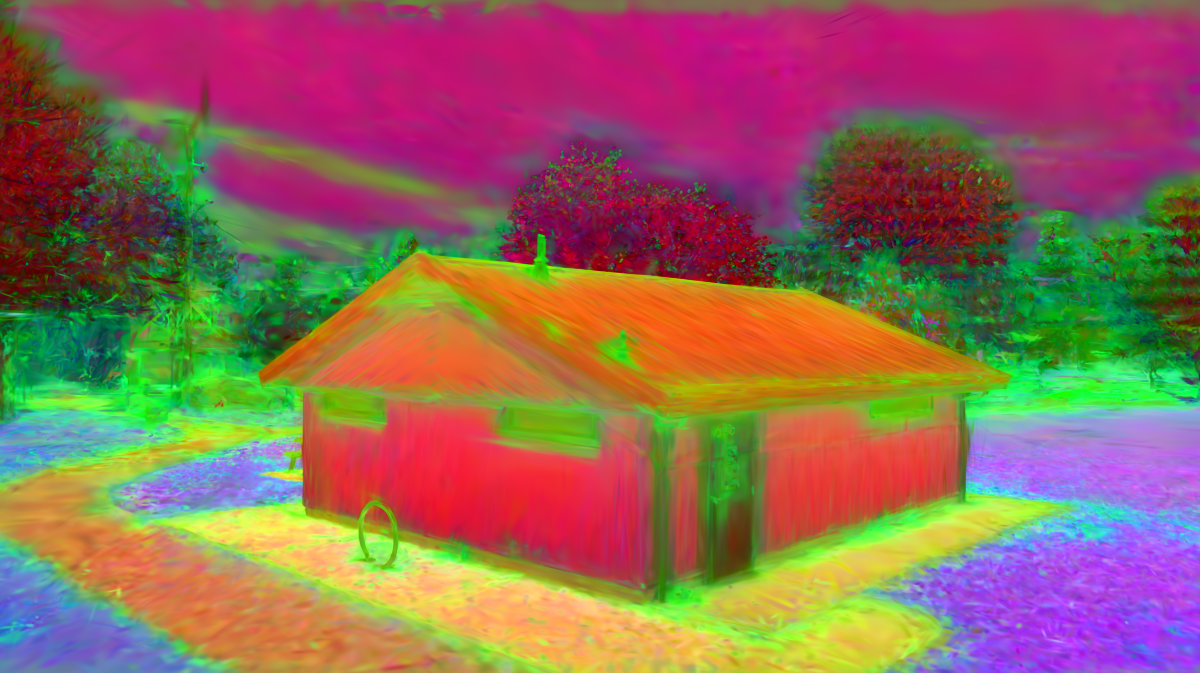}
         \caption{SAM Feature Field}
         \label{FFB-Feature-3DGS-SAM}
     \end{subfigure}
     \hfill
     \begin{subfigure}[b]{0.195\textwidth}
         \centering
         \includegraphics[width=\textwidth]{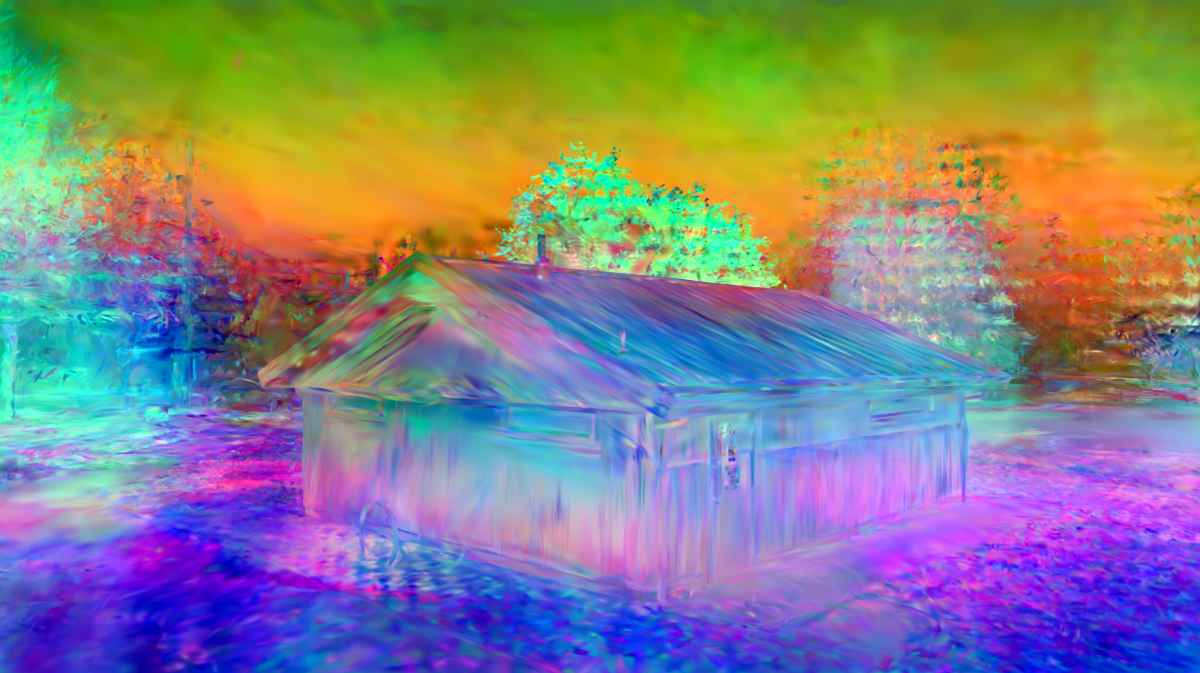}
         \caption{SAM2 Feature Field}
         \label{FFB-Feature-3DGS-SAM2}
     \end{subfigure}
        \caption{Renderings of the 3D feature fields for the Building dataset}
        \label{fig:building-feature-fields}
\end{figure*}

\begin{figure*}
     \centering
     \begin{subfigure}[b]{0.195\textwidth}
         \centering
         \includegraphics[width=\textwidth]{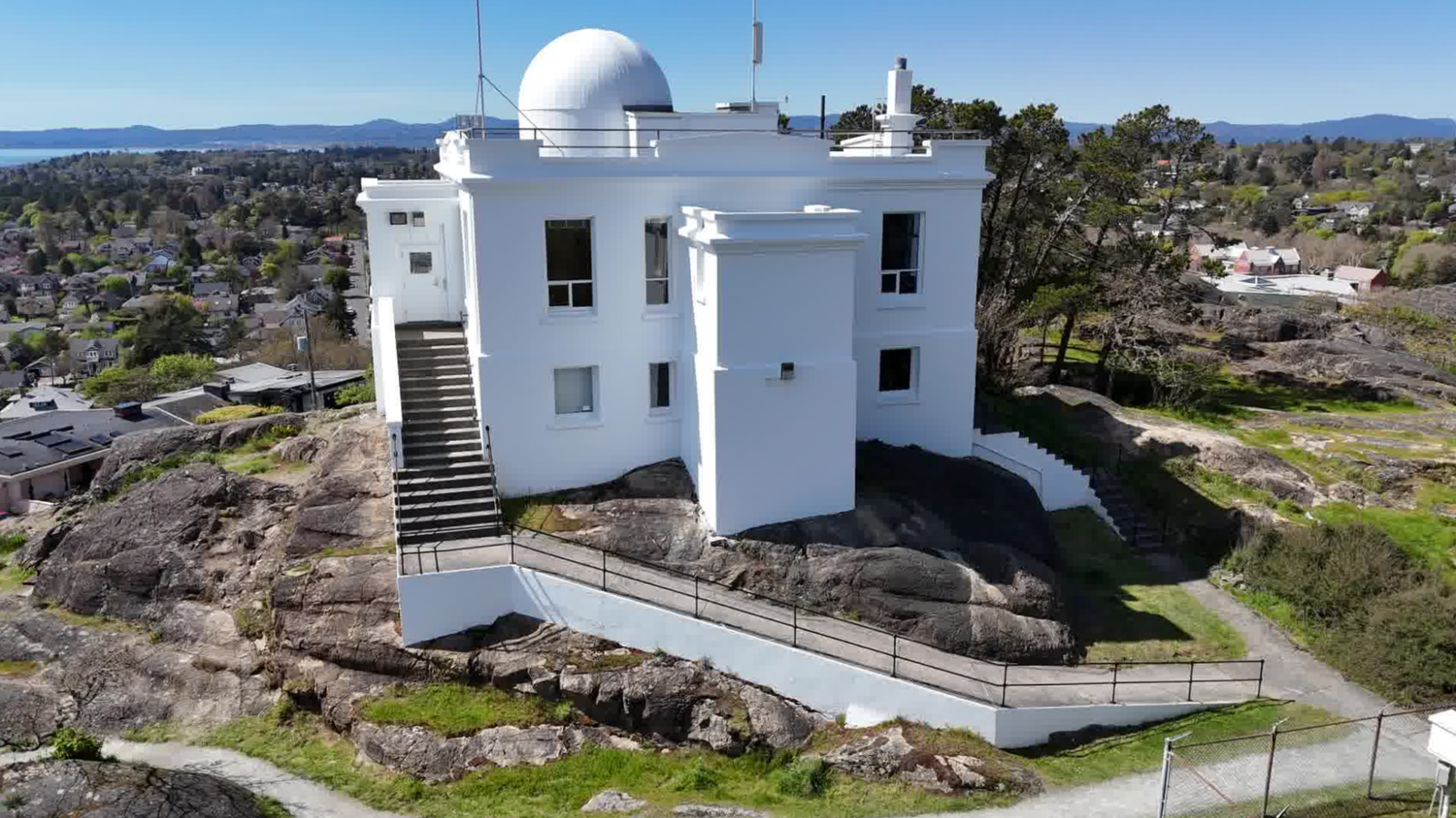}
         \caption{Ground Truth Image}
         \label{FFO-LangSplat-SAM}
     \end{subfigure}
     \hfill
     \begin{subfigure}[b]{0.195\textwidth}
         \centering
         \includegraphics[width=\textwidth]{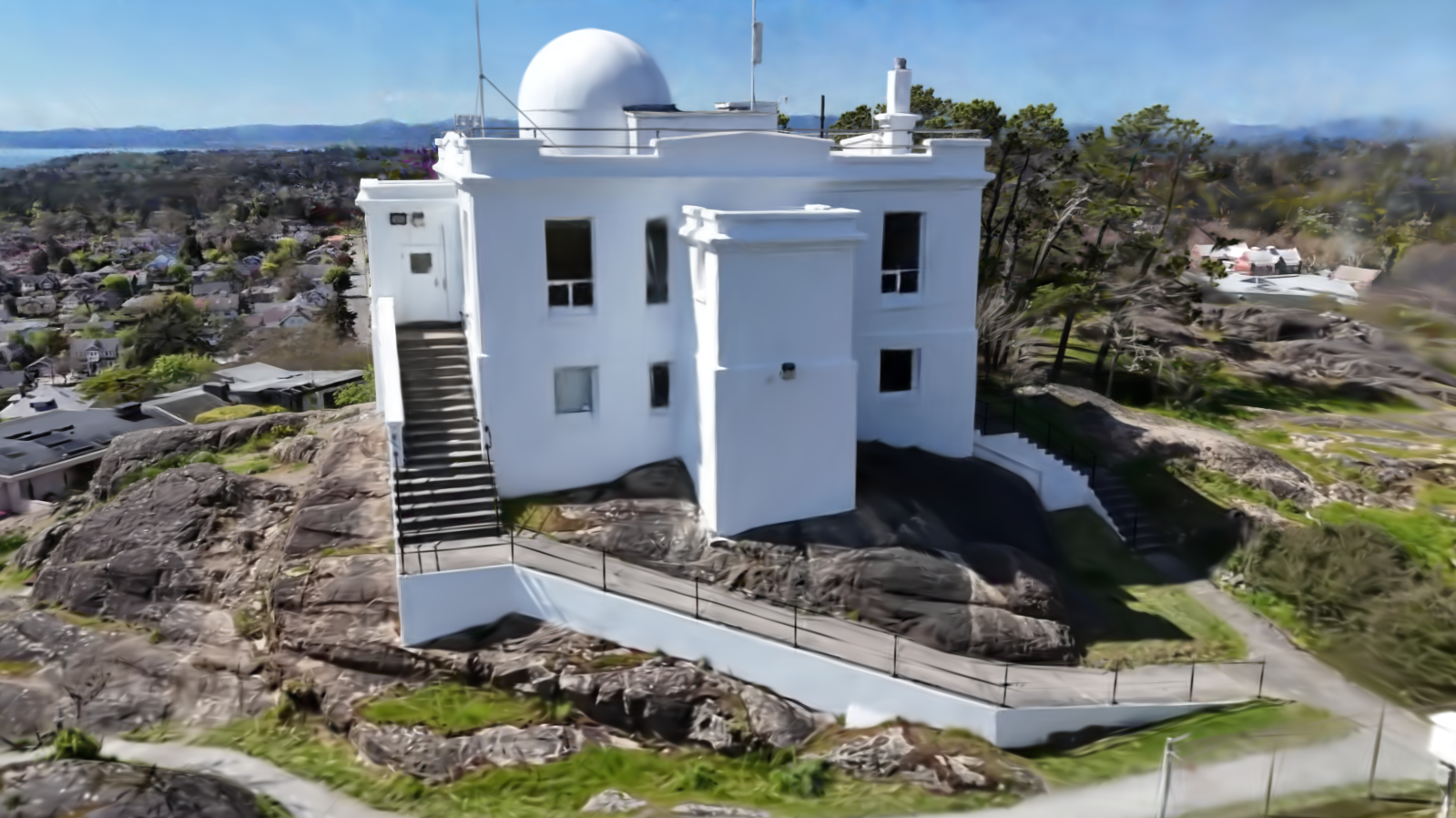}
         \caption{3DGS Rendering}
         \label{FFO-LangSplat-SAM2}
     \end{subfigure}
     \hfill
     \begin{subfigure}[b]{0.195\textwidth}
         \centering
         \includegraphics[width=\textwidth]{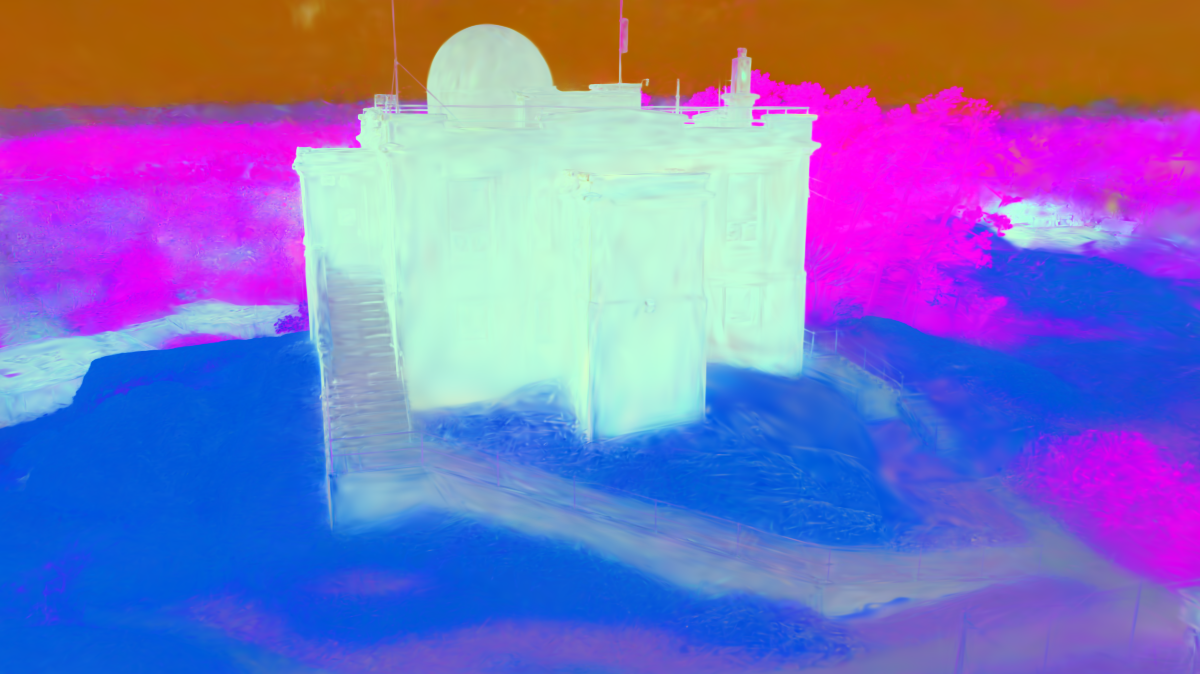}
         \caption{LSeg Feature Field}
         \label{FFO-Feature-3DGS-LSeg}
     \end{subfigure}
     \hfill
     \begin{subfigure}[b]{0.195\textwidth}
         \centering
         \includegraphics[width=\textwidth]{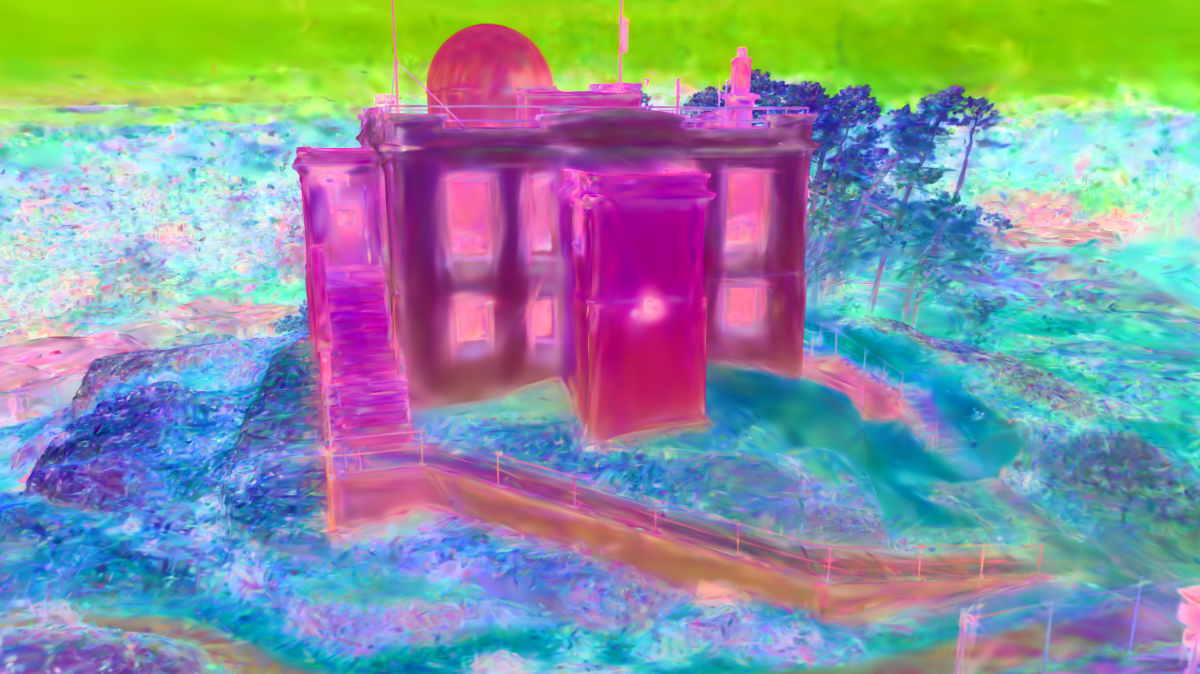}
         \caption{SAM Feature Field}
         \label{FFO-Feature-3DGS-SAM}
     \end{subfigure}
     \hfill
     \begin{subfigure}[b]{0.195\textwidth}
         \centering
         \includegraphics[width=\textwidth]{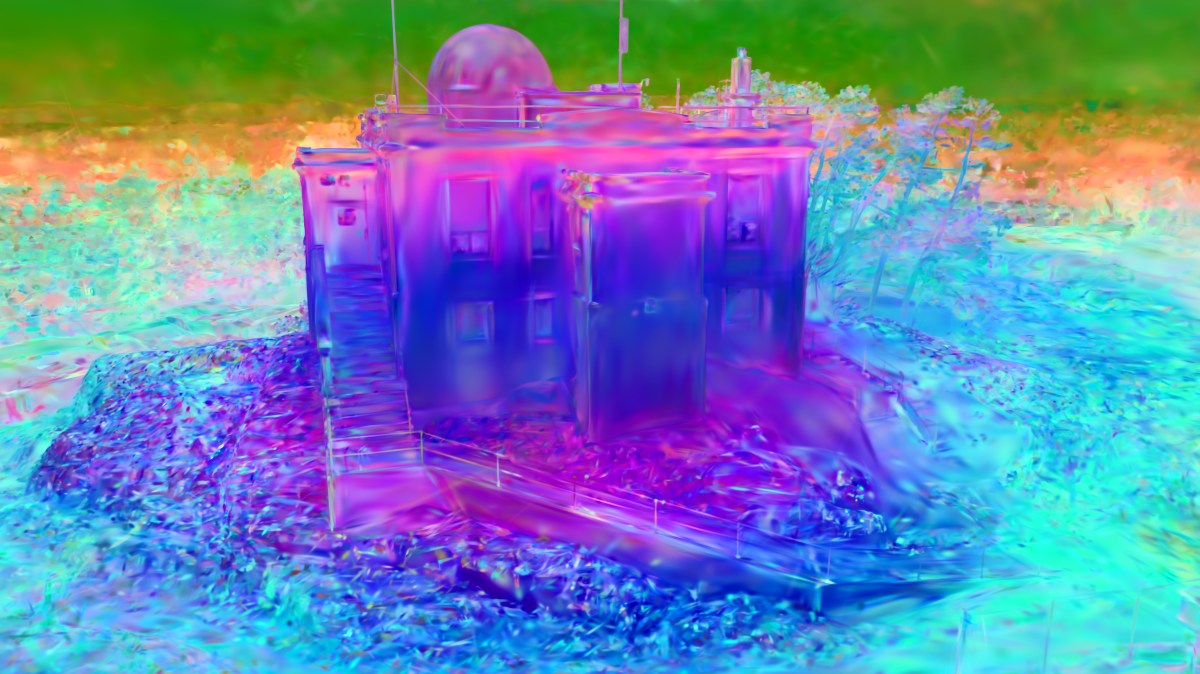}
         \caption{SAM2 Feature Field}
         \label{FFO-Feature-3DGS-SAM2}
     \end{subfigure}
        \caption{Renderings of the 3D feature fields for the Observatory dataset}
        \label{fig:observatory-feature-fields}
        \vspace{-1em}
\end{figure*}

\subsection{3D Feature Field Visualization}

To evaluate the quality and characteristics of different feature field backbones, we trained separate Feature-3DGS models using CLIP-LSeg, SAM, and SAM2 feature extractors. Each model was trained for 7,000 iterations, following the default dual-branch loss in Feature-3DGS. The feature fields were rendered from novel views and compared across both datasets.

\Cref{fig:building-feature-fields} and \Cref{fig:observatory-feature-fields} show qualitative results for the Building and Observatory datasets, respectively. Each row presents a ground truth image, the corresponding 3DGS rendering, and feature fields derived from CLIP-LSeg, SAM, and SAM2 for the same viewpoint.

The CLIP-LSeg feature fields demonstrate strong semantic coherence. Large-scale elements such as the building, trees, and grassy areas appear uniformly colored, reflecting the model’s ability to capture contextual scene understanding. For example, in \Cref{fig:building-feature-fields}(c), nearly all building surfaces share a consistent feature encoding, while vegetation and ground areas are represented distinctly.

The SAM-derived feature fields, shown in \Cref{fig:building-feature-fields}(d) and \Cref{fig:observatory-feature-fields}(d), are noticeably more granular. They produce distinct feature encodings for subcomponents like windows, doors, walls, and roof panels, suggesting stronger object-level discrimination. However, this granularity can also introduce inconsistency in texture-heavy or reflective regions.

In contrast, the SAM2 feature fields (\Cref{fig:building-feature-fields}(e), \Cref{fig:observatory-feature-fields}(e)) appear visually noisier and less interpretable. While SAM2 captures many fine-grained regions, the color distribution lacks clear spatial coherence, and feature clustering does not align well with meaningful object boundaries. This is consistent with SAM2's design trade-offs, which prioritize computational efficiency and real-time inference, especially for video segmentation, over improved performance on static image tasks.

These results suggest that for outdoor aerial scenes, CLIP-LSeg provides strong contextual grouping, while SAM offers more localized semantic detail. SAM2, though efficient, may require domain-specific tuning to yield interpretable feature fields in complex environments.

\subsection{Language-Driven Feature Highlighting}

\begin{figure}
     \centering
     \begin{subfigure}[b]{0.235\textwidth}
         \centering
         \includegraphics[width=\textwidth]{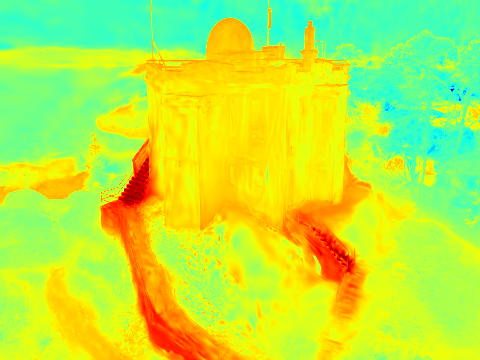}
         \caption{LSeg Logit Heatmap}
         \label{b-lseg-heatmap}
     \end{subfigure}
     \hfill
     \begin{subfigure}[b]{0.235\textwidth}
         \centering
         \includegraphics[width=\textwidth]{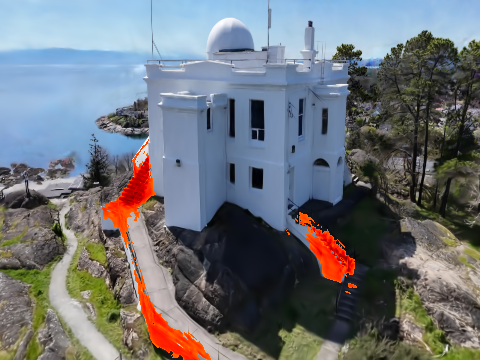}
         \caption{Score-based Segmentation}
         \label{b-lseg-heatmap-overlay}
     \end{subfigure}
        \caption{LSeg thresholded segmentation with the prompt "stairs with metal railing"}
        \label{fig:observatory-lseg}
        \vspace{-1em}
\end{figure}

To enable language interaction with the 3D reconstructions, we use the CLIP-LSeg trained Feature-3DGS model to compute heatmaps based on cosine similarity between CLIP-embedded text prompts and the rendered feature fields. \Cref{fig:observatory-lseg} shows results for the prompt \textit{“stairs with metal railing”}.

\Cref{fig:observatory-lseg}(a) presents the full similarity heatmap rendered from a novel view. The relevant region—metal staircases on the observatory—appears in bright red, with lower-confidence regions shown in darker tones. \Cref{fig:observatory-feature-fields}(b) overlays a thresholded version of this heatmap on the 3DGS rendering, providing a segmentation of the queried region.

Qualitatively, this method produces reasonably accurate semantic localizations for large, distinctive objects. However, it struggles with small, visually ambiguous elements such as windows or signage. Thresholding introduces sensitivity to score distribution and can either oversegment or miss relevant areas, depending on the prompt and view.

While coarse, this step enables language-based identification of scene regions and serves as the first stage in our segmentation pipeline.

In general, prominent and structurally distinct features such as roofs and domes are successfully highlighted in response to relevant prompts. More subtle features, such as individual windows or entranceways, are more challenging to isolate, particularly when surrounding textures are homogeneous or when features are small relative to the overall scene scale. These results suggest that while language-grounded feature fields are promising for high-level semantic querying, additional refinement is needed for fine-grained inspection tasks.

\subsection{Prompted Segmentation via SAM and SAM2}

\begin{figure}
     \centering
     \begin{subfigure}[b]{0.235\textwidth}
         \centering
         \includegraphics[width=\textwidth]{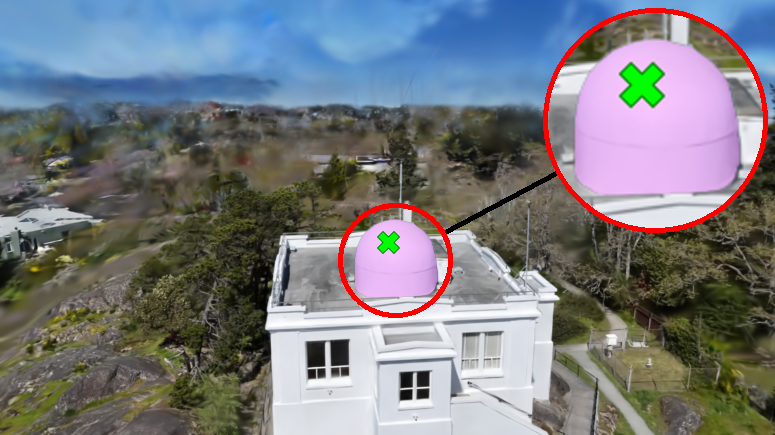}
         \label{o-sam-1}
     \end{subfigure}
     \hfill
     \begin{subfigure}[b]{0.235\textwidth}
         \centering
         \includegraphics[width=\textwidth]{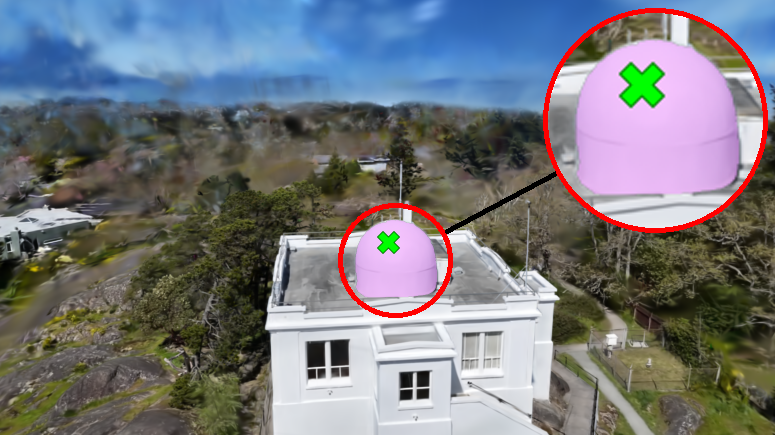}
         \label{o-sam-2}
     \end{subfigure}
     \\[-1em]
     \begin{subfigure}[b]{0.235\textwidth}
         \centering
         \includegraphics[width=\textwidth]{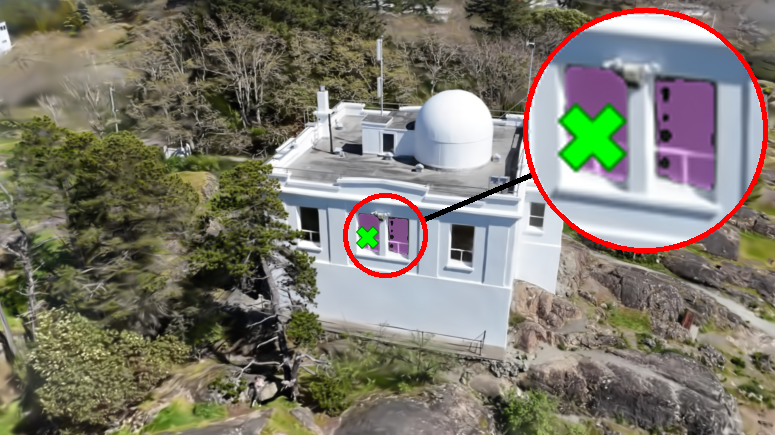}
         \label{0-sam-2}
     \end{subfigure}
     \hfill
     \begin{subfigure}[b]{0.235\textwidth}
         \centering
         \includegraphics[width=\textwidth]{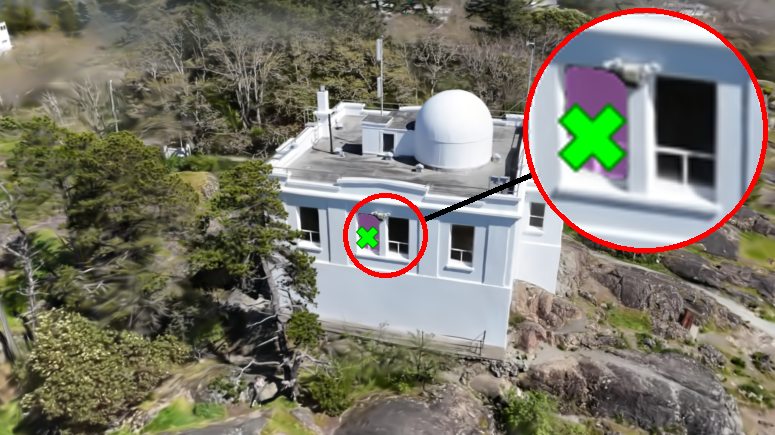}
         \label{o-sam2-2}
     \end{subfigure}
     \\[-1em]
     \begin{subfigure}[b]{0.235\textwidth}
         \centering
         \includegraphics[width=\textwidth]{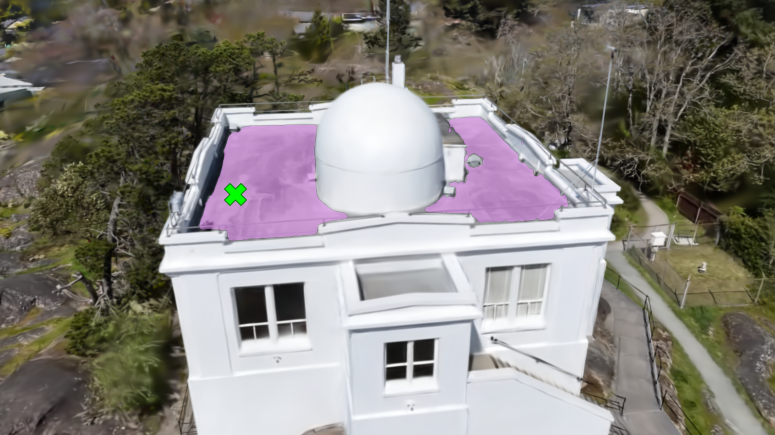}
         \caption{SAM}
         \label{0-sam-3}
     \end{subfigure}
     \hfill
     \begin{subfigure}[b]{0.235\textwidth}
         \centering
         \includegraphics[width=\textwidth]{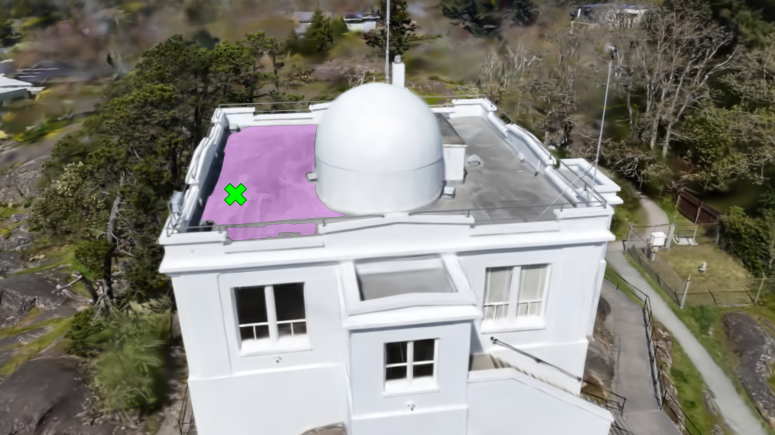}
         \caption{SAM2}
         \label{o-sam2-3}
     \end{subfigure}
        \caption{Comparison of the point-prompted SAM segmentations.}
        \label{fig:SAM-comparison}
        \vspace{-1em}
\end{figure}

To refine segmentations beyond thresholded heatmaps, we use the argmax point from the LSeg similarity map as a prompt to both SAM and SAM2. Segmentation masks are generated from the same viewpoint using each model and compared side-by-side. \Cref{fig:SAM-comparison} shows results for three prompt locations on the Observatory dataset.

In the first example, both SAM and SAM2 segment the targeted window, but SAM outputs a broader region that includes adjacent panes. SAM2 produces a tighter mask around the single pane aligned with the prompt point. The second example shows a segmentation of the observatory roof; SAM segments the entire roof structure, while SAM2 restricts the mask to the roof half nearest to the prompt. This is likely due to a minor occlusion dividing the roof in the rendered view.

In general, SAM consistently produces broader masks that align well with object-level regions, whereas SAM2 returns more focused results. This behavior is likely due to differences in their architecture, particularly the smaller encoder in SAM2 and the inclusion of memory blocks aimed at handling video segmentation tasks. While both models generalize well, their performance differs across modalities, with SAM better suited for static images and SAM2 optimized for video content. Importantly, the differences are relatively consistent across prompts and scenes.

\subsection{End-to-End Language-Guided Segmentation}

\begin{figure}
     \centering
     \begin{subfigure}[b]{0.235\textwidth}
         \centering
         \includegraphics[width=\textwidth]{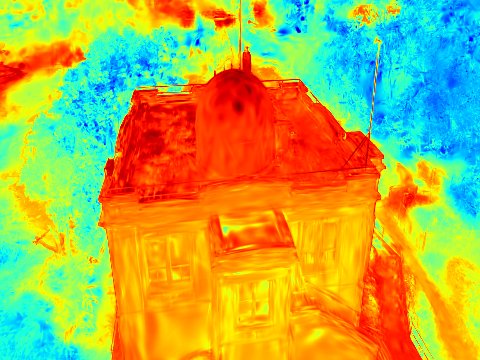}
         \caption{LSeg Logit Heatmap}
     \end{subfigure}
     \hfill
     \begin{subfigure}[b]{0.235\textwidth}
         \centering
         \includegraphics[width=\textwidth]{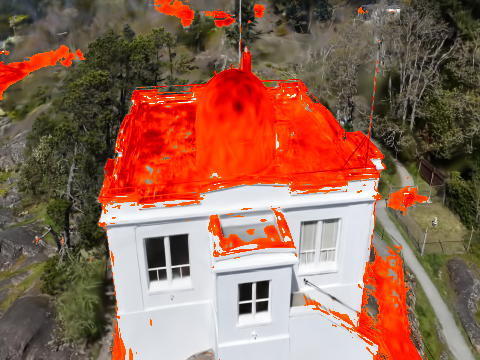}
         \caption{Score-based Segmentation}
     \end{subfigure}
     \hfill
     \begin{subfigure}[b]{0.45\textwidth}
         \centering
         \includegraphics[width=\textwidth]{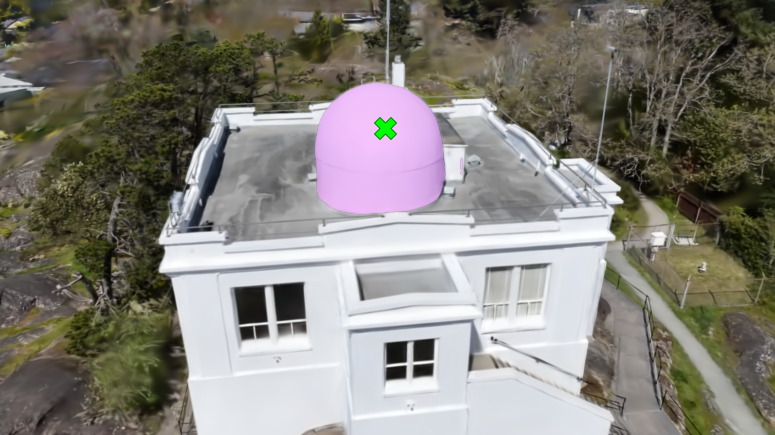}
         \caption{Point-prompted SAM segmentation}
     \end{subfigure}
        \caption{End-to-end language prompted segmentation using LSeg to point-prompt SAM from the text prompt "dome"}
        \label{fig:E2E}
        \vspace{-1em}
\end{figure}

Combining the heatmap-based and point-prompted segmentation steps, we demonstrate an end-to-end language-driven pipeline for 3D segmentation. \Cref{fig:E2E} shows results for the prompt \textit{“dome”} on the Observatory dataset.

\Cref{fig:E2E}(a) shows the cosine similarity heatmap produced by CLIP-LSeg. The dome structure is distinctly highlighted in red, while surrounding structures have lower similarity scores. \Cref{fig:E2E}(b) displays the thresholded binary segmentation. Finally, \Cref{fig:E2E}(c) shows the refined segmentation produced by SAM using the highest-activation point in the heatmap as a prompt.

This result demonstrates the feasibility of using language queries to drive object segmentation in outdoor 3D scenes. The existing data collection and analysis pipeline is designed to facilitate segmentation and the identification of objects within complex scenes. Ongoing development aims to enhance this tool into an automated quantitative photogrammetry system capable of measuring the specified objects.

\subsection{Discussion}

Our results highlight key observations on integrating vision-language models within the Feature-3DGS framework for outdoor aerial inspection.

First, the feature fields derived from CLIP-LSeg, SAM, and SAM2 exhibit distinct characteristics. CLIP-LSeg produces smooth, semantically coherent regions (e.g., facades, vegetation), suited for high-level grouping. SAM fields are more granular, capturing local structures such as windows and edges. SAM2, optimized for speed, often yields noisier and less interpretable features. These trends suggest that while CLIP-LSeg excels at semantic grouping, and SAM for precise object localization, both models may require further adaptation to handle the challenges of complex outdoor scenes.

Second, our two-stage segmentation pipeline enables open-vocabulary querying by combining CLIP-based heatmaps with SAM refinement. This approach yields interpretable segmentations for many prompts but has limitations: heatmaps are sensitive to prompt phrasing, and argmax-based point prompting is simplistic. More robust strategies, such as centroid- or region-aware prompts, could improve reliability.

Finally, applying language-guided segmentation to UAV-captured scenes remains challenging. Lighting variation, occlusion, and scale differences degrade semantic consistency. Foundation models like CLIP and SAM are not trained for aerial data, and their performance in this domain is inconsistent. Nonetheless, our results demonstrate that bridging high-fidelity reconstruction with semantic interaction is feasible, even in complex, real-world settings.

\section{CONCLUSIONS AND FUTURE WORK} \label{sec:conclusion}

This work presents a proof of concept for combining language-guided segmentation with 3D Gaussian Splatting to support semantic analysis of aerial inspection scenes. By integrating CLIP-based features and SAM-based refinement, our pipeline enables interactive, open-vocabulary querying of complex outdoor environments without task-specific fine-tuning.

While effective for coarse structures, current models struggle with fine-grained or domain-specific features such as cracks or wiring. Similarly, the reconstruction fidelity, though visually compelling, remains limited for thin or subtle elements.

Future work will focus on:
\begin{itemize}
    \item \textbf{Specialized models:} Fine-tuning or training segmentation models for aerial inspection targets (e.g., corrosion, joints, cracks), combined with stronger grounding techniques like DINO or open-vocabulary vision-language models.
    \item \textbf{Improved 3D reconstruction:} Incorporating recent advances such as 3D Gaussian Ray Tracing \cite{loccoz20243dgrt} and 3DGUT \cite{wu20253dgut} to handle complex optics and improve rendering accuracy in outdoor settings.
    \item \textbf{True 3D feature fields:} Extending 2D-derived features into volumetric 3D fields to support direct segmentation, detection, and language interaction within 3D space.
    \item \textbf{Deployment:} Optimizing for onboard or real-time inference, improving robustness to environmental variability, and developing tailored datasets for aerial infrastructure inspection.
\end{itemize}

These directions aim to advance toward intelligent inspection systems that combine efficient reconstruction with interactive, multimodal scene understanding.

\addtolength{\textheight}{-12cm}   







\bibliographystyle{ieeetr}
\bibliography{refs}

\end{document}